\documentclass[twoside,english]{elsarticle}
\usepackage{ae,aecompl}
\usepackage[T1]{fontenc}
\usepackage[latin9]{inputenc}
\usepackage{geometry}
\usepackage{array}
\usepackage{float}
\usepackage{multirow}
\usepackage{graphicx}

\makeatletter

\providecommand{\tabularnewline}{\\}

\makeatletter
\def\ps@pprintTitle{%
  \let\@oddhead\@empty
  \let\@evenhead\@empty
  \def\@oddfoot{}
  \let\@evenfoot\@oddfoot
}
\makeatother

\usepackage{fancyhdr}
\usepackage{lineno}

\usepackage{pdflscape}
\usepackage[nolist,nohyperlinks]{acronym}

\makeatother

\usepackage{babel}
\begin{document}
\begin{frontmatter}

\title{{\Large{}The process of macroprudential oversight in Europe}\tnoteref{mytitlenote}}

\tnotetext[mytitlenote]{We are grateful to the 23 persons from 12 institutions of the European
System of Financial Supervisors for participating in the interviews
conducted in this project, as well as to Alistair Milne and Mark D.
Flood for valuable comments and discussion. The paper has also benefited
from comments during presentations at the anonymous participant institutions,
ranging from small-scale meetings to institution-wide seminars, as
well as the Conference on Data Standards, Information and Financial
Stability at Loughborough University, UK on 11--12 April, 2014. Corresponding
author: Peter Sarlin, Goethe University, Center of Excellence SAFE,
Grüneburgplatz 1, 60323 Frankfurt am Main, Germany. E-mail: psarlin@abo.fi.}

\author[P1,P2,P3]{Peter Sarlin}

\author[H]{and Henrik J. Nyman}

\address[P1]{Center of Excellence SAFE at Goethe University Frankfurt, Germany}

\address[P2]{Department of Economics, Hanken School of Economics, Helsinki, Finland}

\address[P3]{RiskLab Finland at Arcada University of Applied Sciences, Helsinki,
Finland}

\address[H]{Institute for Advanced Management Systems Research (IAMSR), Åbo
Akademi University, Turku, Finland}
\begin{abstract}
The 2007--2008 financial crisis has paved the way for the use of macroprudential
policies in supervising the financial system as a whole. This paper
views macroprudential oversight in Europe as a process, a sequence
of activities with the ultimate aim of safeguarding financial stability.
To conceptualize a process in this context, we introduce the notion
of a public collaborative process (PCP). PCPs involve multiple organizations
with a common objective, where a number of dispersed organizations
cooperate under various unstructured forms and take a collaborative
approach to reaching the final goal. We argue that PCPs can and should
essentially be managed using the tools and practices common for business
processes. To this end, we conduct an assessment of process readiness
for macroprudential oversight in Europe. Based upon interviews with
key European policymakers and supervisors, we provide an analysis
model to assess the maturity of five process enablers for macroprudential
oversight. With the results of our analysis, we give clear recommendations
on the areas that need further attention when macroprudential oversight
is being developed, in addition to providing a general purpose framework
for monitoring the impact of improvement efforts.\end{abstract}
\end{frontmatter}

\newpage{}

\textit{\footnotesize{}\hspace{3cm}\textquotedblleft }\emph{\footnotesize{}Putting
in place this emerging framework for financial stability will be difficult.
As we have }{\footnotesize \par}

\textit{\footnotesize{}\hspace{3cm}}\emph{\footnotesize{}seen, it
requires new consensus and tools, but it also requires additional
resources, better}{\footnotesize \par}

\textit{\footnotesize{}\hspace{3cm}}\emph{\footnotesize{}data, time,
and international cooperation to enable us to take timely action.'' }{\footnotesize \par}

\textit{\footnotesize{}\hspace{3cm}}{\footnotesize{}-- Jaime Caruana,
General Manager of the BIS, Washington DC, 23/04/2010}{\footnotesize \par}

\section{Introduction}

The current financial crisis has highlighted the importance of a system-wide,
or macroprudential, approach to safeguarding financial stability (e.g.,
\citet{Borio2011}). Rather than being only concerned with the stability
of individual financial institutions, it involves the tasks of preventing
and mitigating systemic risks to avoid widespread financial distress
and at the same time also consider economic growth (e.g., ESRB Regulation,
Article 3(1)). Given these aims, macroprudential oversight can be
viewed as a process, which exhibits inherent complexity. In the case
of macroprudential oversight in Europe, the system of financial supervisors
include a large number of actors both at the European and national
level, with a common goal of safeguarding financial stability. With
the European System of Financial Supervisors currently in the making
(e.g., \citet{EU2009a,EU2009b}), now is the time to understand activities,
define responsible entities and discern how the actors interact. Ongoing
public and academic discussion has targeted individual aspects, which
oftentimes concern roles of individual institutions, such as legal
challenges (e.g., UK vs. ESMA and Germany vs. banking union), the
mandate within central banks or politicians (\citet{Issing2009}),
the role of central banks within the ESRB (\citet{Goodhart2011}),
ECB as a lender of last resort in the government bond market \citep{deGrauwe2013},
and a role of E(S)CB in euro area vs. ESRB in Europe (\citet{schoenmaker2013}).
Despite these important discussions, little consideration has been
given to the inner workings and broad underpinnings of macroprudential
oversight at large. In this vein, this paper takes a comprehensive
approach to assessing macroprudential oversight in Europe through
a process perspective. 

Process management concerns the analysis and improvement of ways of
working through systematic representations of organizations' processes
(e.g., \citet{Segatt0etal2013}). This provides means not only for
representing current ways of working, but also for reaching consensus
on what form the process should take and possible areas of improvement
(e.g., \citet{Hammer2010} and \citet{Eikebrokketal2011}). Yet, this
describes the functioning and optimization of processes from the viewpoint
of a single organization. In the public sector, processes may be of
a different nature. They may involve multiple organizations with a
common objective, where the actors take a collaborative approach to
reaching the final goal. We denote these as public collaborative processes
(PCPs). PCPs involve a number of dispersed organizations cooperating
under various unstructured forms. For instance, public health may
involve activities ranging from nutritional health advice to the services
of both private and public health care centers and hospitals. While
being of different nature, in this paper we argue that PCPs can and
should still be examined through the lens of process management.

Macroprudential oversight has been described as a process comprising
the following high-level tasks (e.g., \citet{ECB2010}): risk identification,
risk assessment and risk communication, as well as the assessment
and implementation of policies. While being relatively well-defined
at this level, the tasks still involve a number of uncertainties and
limitations that challenge the functioning of the process. At a more
detailed level, the process is among other things characterized by
a large number of involved actors, a national and supranational level,
an inherent political dimension, vast amounts of data, and decision-making
based upon expert judgment and numerical methods. Further, macroprudential
oversight exhibits a high degree of dependence between and within
tasks as in any process. Given the aim of process management, in which
activities are to be coordinated so that disperse tasks performed
by many different partners act as one seamless process, concerns can
be raised about the functioning of the European macroprudential oversight
process (as noted in the first paragraph). Understanding dependence
within the tasks becomes particularly crucial when moving towards
a more detailed representation of the process, which is the essence
of process management. However, an essential prerequisite for detailing
and documenting a process is a common agreement of the activities
at the focus of attention and the roles of all involved parties, i.e.,
who is doing what. As a prerequisite for process management, this
paper examines the maturity of the macroprudential oversight process
through various constructs, such as process design, metrics, and ownership.
As such, by identifying areas of improvement and directing attention
at needs, this work lays the foundation for reaching a more mature
process for safeguarding financial stability. 

With the aim of assessing the maturity of macroprudential oversight
in Europe, this paper conducts an interview study and builds an analysis
model. The study includes interviews with 23 key policymakers and
supervisors from 12 organizations involved in macroprudential oversight
across Europe. The interviewees were asked a series of questions relating
to macroprudential oversight in Europe, which were mapped to five
so-called process enablers (\citet{Hammer2007,Hammer2010}). The answers
were used to assess the maturity of macroprudential oversight in relation
to (\emph{i}) design, (\emph{ii}) metrics, (\emph{iii}) performers,
(\emph{iv}) infrastructure, and (\emph{v}) ownership of the process.
We provide an analysis model for assessing the maturity of the five
process enablers for macroprudential oversight in Europe, in which
we measure both the level and dissensus in process readiness. This
provides a basis to give clear recommendations regarding the areas
that need further attention when developing macroprudential oversight
in Europe. Rather than an ending point, our analysis model also puts
forward a structured approach not only to assess the current state
but also to monitor the impact of any improvement efforts. As such,
a further aim of this study is to inspire other assessments with a
similar goal in the future. 

The paper is structured as follows: first, we briefly discuss macroprudential
oversight in light of currently available material. This is followed
by a discussion of business process management, and how this can be
applied in the context of PCPs. In this section, we also examine the
differences between standard processes and PCPs, and show that despite
differences a similar approach to formalizing and improving ways of
working is appropriate. In the next section, we discuss our approach
to analyzing data and the results of our analysis of process readiness,
and present the areas of improvement needed for reaching a mature
process for safeguarding financial stability. In the conclusion we
put forward an agenda for future research, suggesting how we can move
from an analysis of process readiness to full-blown process management.

\section{Macroprudential oversight}

Paraphrasing Milton Friedman's statement about Keynesians, \citet{Borio2011}
stated \emph{\textquotedblleft We are all macroprudentialists now.\textquotedblright{}}
The 2007--2008 financial crisis has paved the way for the macroprudential
approach to safeguarding financial stability, which has now grown
consensus among the academic and policymaking communities alike. Yet,
it is no new concept. The BIS applied the term to describe a system-wide
orientation of regulatory frameworks already in the 1970s (see, e.g.,
\citet{BIS1986}). The series of recently established bodies for macroprudential
supervision, as well as their effective or planned mandates, obviously
also motivates understanding and disentangling their specific tasks
and functions, such as the European Systemic Risk Board in Europe,
the Financial Policy Committee in the UK, and the Financial Stability
Oversight Council in the US.

\subsection{Market imperfections and systemic risk }

A comprehensive macroprudential approach to safeguarding financial
stability obviously starts from a thorough understanding of the inner
functioning, particularly potential dysfunctioning, of the financial
system. While definitions related to financial stability remain to
be disputed in the literature, one notion that few oppose is that
a key aim is to have a resilient and well-functioning financial system.
One characterization of such a financial system is through the three
pillars proposed by \citet{FellSchinasi2005}: well-managed financial
institutions, efficiently functioning financial markets and a strong
and robust financial infrastructure. That said, the frequent incidences
of costly financial crises do, however, indicate that the three pillars
do have defects. While each recurrence of financial instability may
have sources of its own kind, market imperfections like asymmetric
and incomplete information, externalities and public-good characteristics
and incomplete markets can be a threat to the functioning of the financial
system. These imperfections, when being related to a financial sector,
may lead to significant fragility of not only individual institutions,
but also the entire system, as noted for instance by \citet{Carletti2008}.
While not being directly caused by market imperfections, \citet{deBandtHartmann2002}
relate fragilities in financial systems to three causes: (\emph{i})
the structure of banks, (\emph{ii}) the interconnection of financial
intermediaries, and (\emph{iii}) the information intensity of financial
contracts. The material risks of these fragilities support the role
of governments and other supervisory authorities in addressing and
monitoring financial instability, which also points to financial stability
being a common good and systemic risk an externality. 

Beyond market imperfections, we can concretize the fragility of financial
systems through the notion of systemic risk. Herein, we follow the
definition of three forms of systemic risk by \citet{deBandt2009}:
(\emph{i}) endogenous build-up and unraveling of widespread imbalances;
(\emph{ii}) exogenous aggregate shocks; and (\emph{iii}) contagion
and spillover. The first form of systemic risk focuses on the unraveling
of widespread imbalances and is illustrated by the presence of risks,
vulnerabilities and imbalances in banking systems and the overall
macro-financial environment prior to financial crises. Early and later
literature alike have identified common patterns in underlying vulnerabilities
preceding financial crises (e.g., \citet{Minsky1982} and \citet{ReinhartRogoff2008}).
The second type of systemic risk, exogenous aggregate shocks, have
been shown to co-occur with financial instabilities (e.g., \citet{Gorton1988}
and \citet{DemirgucDetragiache1998}). One example is the collapse
of banks during recessions due to the vulnerability to economic downturns.
The contagion literature provides evidence on the final, third form
of systemic risk, which involves the cross-sectional transmission
of financial instability (e.g., \citet{UpperWorms2004} and \citet{LelyveldLiedorp2006}).
Here, episodes of financial instabilities have been shown to relate
to the failure of one financial intermediary causing the failure of
another.

\subsection{The macroprudential oversight process}

The above described market imperfections, and thereby caused systemic
risks, are a premise for macroprudential oversight. Accordingly, an
essential task is the aim of signaling these systemic risks at an
early stage. This necessitates access to a broad toolbox of approaches
to measure and analyze system-wide threats to financial stability.
Broadly speaking, tools and models can be divided into those for early
identification and assessment of systemic risks. \citet{ECB2010}
provides a mapping of tools to the above listed three forms of systemic
risk: (\emph{i}) early-warning indicators and models, (\emph{ii})
macro stress-testing models, and (\emph{iii}) contagion models. First,
by focusing on the presence of vulnerabilities and imbalances in an
economy, early-warning models can be used to derive probabilities
of being in a vulnerable state, in which a shock descending from any
source may trigger a systemic financial crises (e.g., \citet{Alessi2011520}
and \citet{Duca2012}). Second, macro stress-testing models provide
means to assess the resilience of the financial system to a wide variety
of aggregate shocks, such as economic downturns (e.g., \citet{Castrenetal2009}
and \citet{Hirtleetal2009}). Third, contagion and spillover models
can be employed to assess how resilient the financial system is to
cross-sectional transmission of financial instability (e.g., \citet{IMF}).
In addition, the literature has also provided a large set of coincident
indicators to measure the contemporaneous level of systemic risk (e.g.,
\citet{Holloetal2012}). While coincident measures may be used to
identify, signal and report on heightened stress, they are not designed
for early identification and assessment of risk. 

\begin{figure}[H]
\begin{centering}
\includegraphics[width=1\columnwidth]{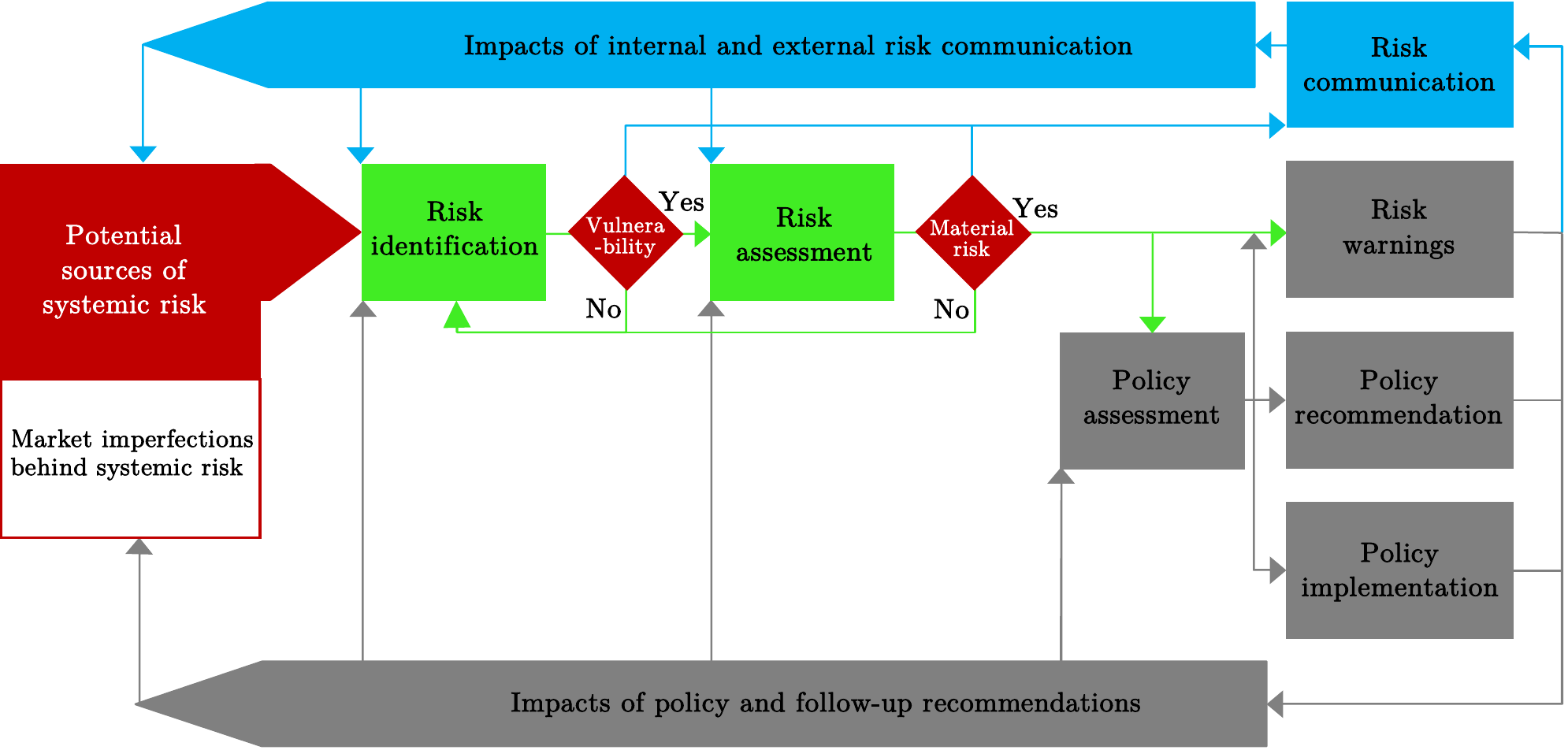}
\par\end{centering}

\textbf{\scriptsize{}Notes}{\scriptsize{}: The figure illustrates
the macroprudential oversight process. The red components represent
risks and vulnerabilities and the green components represent the need
for risk identification and assessment, gray components represent
policy assessment, risk warnings, policy recommendations and policy
implementations, and blue components represent risk communication.
The figure is an adapted version of that in \citet{ECB2010} and \citet{Sarlin2013SWIFT}.}{\scriptsize \par}

\centering{}\protect\caption{\label{fig:The-macroprudential-oversight}The macroprudential oversight
process}
\end{figure}

Despite the importance of analysis, in which risk identification and
assessment is in focus, the most central part of macroprudential oversight
relates to policy interventions. In terms of a process, Figure \ref{fig:The-macroprudential-oversight}
puts forward the steps in the process that a macroprudential supervisory
body follows. As described by \citet{ECB2010}, macroprudential oversight
can be related to three steps: (\emph{i}) risk identification, (\emph{ii})
risk assessment, and (\emph{iii}) risk communication, policy assessment,
risk warnings and policy recommendations and implementation. The process
in Figure \ref{fig:The-macroprudential-oversight} deviates from \citet{ECB2010}
by disentangling the final step into two separate feedback loops,
as proposed by \citet{Sarlin2013SWIFT}. In the figure, red components
represent risks and vulnerabilities, green components represent the
need for risk identification and assessment, gray components represent
policy assessment, risk warnings, policy recommendations and policy
implementations, and blue components represent risk communication.
With no detailed treatment, we present herein the key tasks and tools
used in each step.

In the \emph{first }step of the supervisory process, the key focus
is on identifying risks to stability and potential sources of vulnerability.
The vulnerabilities and risks could exist in any of the three pillars
of the financial system: financial intermediaries, financial markets
and financial infrastructure. The necessary analytical tools to identify
possible risks, vulnerabilities and triggers come from the set of
early-warning models and indicators, combined with the use of market
intelligence, and expert judgment and experience. This involves ranking
risks and vulnerabilities as per intensity, as well as for assigning
probabilities to specific shocks or future systemic events. 

In the \emph{second }step of the process, the rankings and probabilities
may be used to assess the identified risks. Beyond market intelligence,
as well as expert judgment and experience, risk assessment makes use
of analytical tools mainly from the set of macro stress-testing models
and contagion models. In macro stress-testing, simulations of most
plausible risk scenarios show the degree of impact severity on the
overall financial system, as well as its components. Contagion models,
on the other hand, might be used through counterfactual simulations
to assess the impact of specific failures on the entire financial
system and individual institutions. The first and the second step
of the process should not only provide a list of risks ordered according
to possible severity, but also contain their materialization probabilities,
losses given their materialization, and real losses in output and
welfare, as well as their possible systemic impact. Hence, these two
initial steps in the process aim at early risk identification and
assessment and steer subsequent actions for safeguarding financial
stability. 

The \emph{third }step of the process involves the assessment, recommendation
and implementation of policy actions as early preventive measures,
as well as the communication of risks and vulnerabilities. Based upon
the identified and assessed risks, a macroprudential supervisory body
can consider giving a wide variety of risk warnings and recommendations
for other parties to use policy instruments, as well as implementations
of policies given the instruments at hand. With no detailed discussion
of policies, the most prominent tools to steer system-wide risks include
countercyclical capital buffers (CCBs), and loan-to-value (LTV) and
debt-to-income caps, among many other discussed policy tools. To steer
their decisions, the policy assessment step can make use of the same
analytical tools used for risk identification and assessment. Likewise,
risk warnings and policy recommendations can make use of the analytical
tools. While the use of policy tools may be beyond the mandate of
some macroprudential supervisory bodies, actions tailored to the needs
of a system-wide orientation are obviously a key part of macroprudential
regulation and supervision. As illustrated in Figure \ref{fig:The-macroprudential-oversight},
policies and their recommendations have an impact, not only on the
assessment of policy and identification and assessment of risks, but
obviously also directly on market imperfections and the accumulation
of systemic risks. The feedback of risk communication can, likewise,
be divided into those that affect the risk identification and assessment
and those affecting the financial market at large, where the former
can be targeted with internal communication and the latter with external
communication. Moreover, the information to be communicated might
derive from the risk identification and assessment steps, as well
as from the other tasks in step three.

\subsection{European System of Financial Supervisors}

The European transition towards a common framework for macroprudential
oversight can be motivated with goal of reaching financial stability
in an environment of cross-border finance. Accordingly, paraphrasing
the classic trilemma of monetary policy, the financial trilemma questions
the possibility of simultaneous (\emph{i}) financial stability, (\emph{ii})
financial integration and (\emph{iii}) national financial policies.
As proposed already by \citet{Thygesen2003} and \citet{Schoenmaker2005},
and formalized in \citet{Schoenmaker2011}, one of the three objectives
has to give. In a world of increasing financial and economic integration,
this indicates that the task of producing the public good of financial
stability ought to also involve a supranational level, as proposed
by \citet{deLarosiere2009}.

To understand the set-up of macroprudential oversight in the European
case, we need to focus on the entire system of financial supervisors.
In this section, we briefly discuss the actors and their roles in
the European System of Financial Supervisors (ESFS). Without including
strictly political institutions, the actors involved in safeguarding
financial stability are the following: European Central Bank (ECB),
European Systemic Risk Board (ESRB), national central banks (NCBs),
European Commission (EC), European Banking Authority (EBA), European
Insurance \& Occupational Pensions Authority (EIOPA), European Securities
\& Market Authority (ESMA), and national banking, insurance and securities
supervisors. While their common aim is safeguarding financial stability
at various levels, it is needless to say that disentangling activities
and functions, as well as their connectors, at the level of responsible
entities substantially increases complexity.

\begin{figure}[H]
\begin{centering}
\includegraphics[width=1\columnwidth]{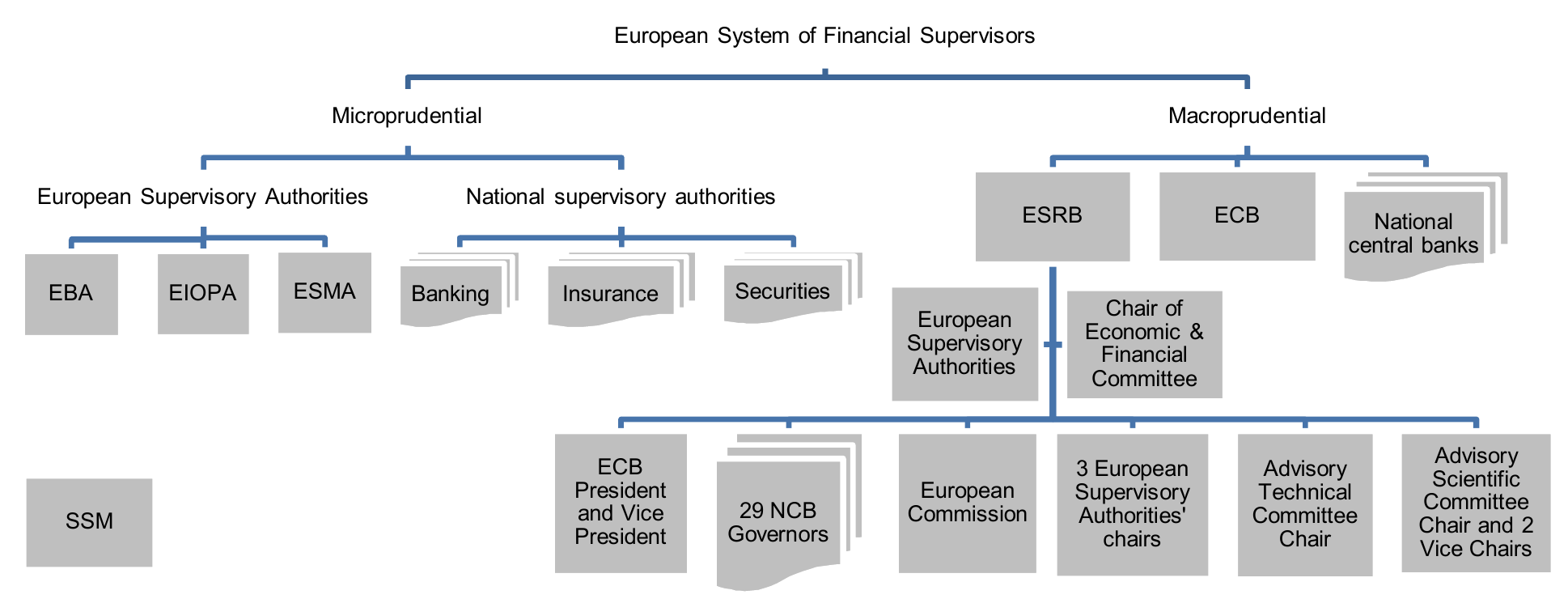}
\par\end{centering}

\textbf{\scriptsize{}Notes}{\scriptsize{}: The figure describes the
structure of the ESFS depicted by the \citet{EU2009a,EU2009b}. It
is adapted from a version in \citet{Hartmann2013}.}{\scriptsize \par}

\centering{}\protect\caption{\label{fig:ESFS}The European System of Financial Supervisors}
\end{figure}

Following the description by the European Commission \citep{EU2009a,EU2009b},
as well as the illustration by \citet{Hartmann2013}, we can move
towards an understanding of the overall division of aims and focus
areas. First, Figure \ref{fig:ESFS} shows that supervision is divided
into microprudential and macroprudential supervisory bodies. Second,
within each of these two branches, the actors consist of both national
and European institutions. Third, there exists a large share of interaction
and collaboration both between and within these two groups of supervisors.
For instance, while microprudential supervisory bodies do not have
macroprudential oversight as their key aim, they are oftentimes the
ones implementing policies according to recommendations by macroprudential
supervisory bodies. Likewise, microprudential information is an important
input to analysis of system-wide risks by macroprudential supervisors.
Yet, while we herein illustrate the complexity of macroprudential
oversight in Europe, we do not provide a detailed description of tasks,
interaction and collaboration.

Moving beyond the stylized representation in Figure \ref{fig:ESFS},
we should not forget that institutional models in Europe significantly
vary among countries. Following the 2012 ESRB recommendation (ref.
ESRB/2011/3) and the EU Capital Requirements Regulation (CRR, Regulation
No. 575/2013), European Union member states were required to set up
a designated authority for macroprudential supervision. In addition
to political and legal heterogeneity across countries, the organizations
mandated with macroprudential oversight include Ministries of Finance,
central banks, financial supervisory authority (FSA) and joint committees
of all above. As is shown in Table \ref{tab:ESFS's-institutional-models},
which is based upon ESRB IWG WP/2013/011 and \citet{Schoenmaker2014},
the central bank is mostly tasked with macroprudential oversight,
yet not at all always.%
\footnote{See \citet{ESRB2014} for a breakdown at the country level.%
} Despite several central banks are mandated with tasks in both monetary
policy and financial supervision, they oftentimes separate micro-
and macroprudential tasks into separate departments, whereas some
FSAs are alone mandated with macroprudential policy. This points also
to another dimension of complexity in the ESFS: multiple policy goals.

\begin{table}[H]
\protect\caption{\label{tab:ESFS's-institutional-models}ESFS's institutional models
for macroprudential oversight}

\begin{centering}
\begin{tabular}{ccccc}
 & {\small{}Model 1} & {\small{}Model 2} & {\small{}Model 3} & {\small{}Model 4}\tabularnewline
\hline 
\hline 
{\small{}Agency } & {\small{}Ministry of Finance} & {\small{}Central bank} & {\small{}FSA} & {\small{}Committee}\tabularnewline
{\small{}Euro area } & {\small{}0 (0\%) } & {\small{}11 (58\%) } & {\small{}4 (21\%)} & {\small{}4 (21\%) }\tabularnewline
{\small{}Non-euro area } & {\small{}1 (10\%)} & {\small{}6 (60\%)} & {\small{}1 (10\%) } & {\small{}2 (20\%) }\tabularnewline
{\small{}Total } & {\small{}1 (3\%) } & {\small{}17 (59\%)} & {\small{}5 (17\%) } & {\small{}6 (21\%)}\tabularnewline
 &  &  &  & \tabularnewline
\end{tabular}
\par\end{centering}

\textbf{\scriptsize{}Notes}{\scriptsize{}: The figure describes the
institutional models in ESFS as reported in reported in ESRB IWG WP/2013/011
and \citet{Schoenmaker2014}. A country breakdown can be found in
\citet{ESRB2014}.}
\end{table}

Moving from financial stability to other policy objectives highlights
the fact that we should not consider one sole process separately but
rather in conjunction with other tasks of the actors in the ESFS.
In the vein of Tinbergen's \citep{Tinbergen1952} analysis, each policy
objective ought to be mapped to a policy instrument, and ideally each
objective has one independent instrument. Following the analysis by
\citet{Schoenmaker2013b}, Table \ref{tab:Synergies-and-conflicts}
shows the potential for synergies and/or conflicts among policy objectives,
when considering monetary, macroprudential and microprudential policies,
objectives and goals together. While each policy has a direct impact
on its own objective, they have a secondary impact on the neighboring
objectives, which all might take the form of synergies and conflicts.
Even though fulfillment of one policy goal may support the fulfillment
of another, the roads to them may involve conflicts. Without weighting
synergies vis-à-vis conflicts, this still highlights the fact that
no one task can be considered in isolation of others. In fact, this
highlights the existence of three separate, yet interacting, processes,
which all ought to be managed given their ultimate goals. 

\begin{table}[H]
\protect\caption{\label{tab:Synergies-and-conflicts}Synergies and conflicts between
policy objectives}

\begin{centering}
\begin{tabular}{ccccc}
{\small{}Policy (typical instruments) } &  & {\small{}Objective} &  & {\small{}Ultimate goal (impact level)}\tabularnewline
\hline 
\hline 
{\small{}Monetary policy} & $\rightarrow$ & \multirow{2}{*}{{\small{}Price stability}} &  & \tabularnewline
{\small{}(short-term interest rate)} & {\small{}$\searrow$} &  & {\small{}$\searrow$} & {\small{}Stable and non-inflationary}\tabularnewline
\cline{1-3} 
{\small{}Macroprudential} & $\nearrow$ &  & $\nearrow$ & {\small{}growth (economic system)}\tabularnewline
{\small{}(LTVs, CCBs)} & $\rightarrow$ & {\small{}Financial stability} &  & \tabularnewline
 & $\searrow$ &  &  & \tabularnewline
\hline 
{\small{}Microprudential} & $\nearrow$ & {\small{}Soundness of individual} & \multirow{2}{*}{{\small{}$\rightarrow$}} & {\small{}Protection of consumers}\tabularnewline
{\small{}(LTVs, capital ratios)} & $\rightarrow$ & {\small{}financial institutions} &  & {\small{}(individual institutions)}\tabularnewline
\hline 
 &  &  &  & \tabularnewline
\end{tabular}
\par\end{centering}

\textbf{\scriptsize{}Notes}{\scriptsize{}: The figure describes the
institutional models in ESFS as reported in reported in ESRB IWG WP/2013/011.
A country breakdown can be found in \citet{ESRB2014}.}
\end{table}

\section{Process management}

This section discusses process management as it is applied in our
context. With a focus on assessing process readiness, we particularly
describe the analysis of public collaborative processes.

\subsection{Business process management }

Broadly speaking, an organizational process can be defined as a chain
of activities that make use of an input to produce an output. This
implies that any organization, public or private, has processes. The
raison d'être for documenting and understanding organizational processes
on the other hand goes back to performance. This can imply that there
is a desire for improvement, or a desire to overcome shortcomings
in performance. Absent any common understanding of a process, it becomes
difficult or impossible to meet or exceed any performance targets.
While a common understanding of how a process works (and how the activities
therein interact) can be implicitly understood by those working on
the task at hand, more complex processes that span across organizational
entities create a need for more formalized communication. 

The roots of quality control and process improvement date back to
the seminal book published in 1911 by \citet{Taylor1911}, which has
hitherto influenced many in the process movement. Aiming at process
performance improvement, business process management (BPM) descends
from two antecedent fields: statistical process control (\citet{Deming1953})
and business process reengineering (\citet{Hammer1990}). BPM as a
field of study takes a holistic approach to managing an organization's
processes. \citet{Segatt0etal2013} define BPM as a discipline focusing
on gaining a common understanding of processes with the aim of continuously
seeking improvement through a feedback cycle, while at the same time
also aligning these processes to organizational strategies. They divide
BPM to six distinct phases (adapted from \citet{ABPMP2009}): 
\begin{enumerate}
\item \emph{Planning}: In this initial stage, executive sponsors, roles
and responsibilities, goals and overall purpose of the BPM exercise
are determined. 
\item \emph{Analysis}: This involves understanding of the current state
of organizational processes. 
\item \emph{Design and modeling}: At this stage, a more detailed representation
of as-is and/or to-be processes are created. Essentially, this phase
focuses on gaining answers to questions such as what is done, by whom,
when, how, and in which organizational entity. This is also referred
to as process modeling, whereby an external representation of the
process is documented as a process model ( \citet{Eikebrokketal2011}). 
\item \emph{Implementation}: Here, activities in the organization are, if
necessary, adapted to findings from previous stages. 
\item \emph{Monitoring and control}: Here, performance metrics are analyzed
to see whether these give raise to further changes in the organization. 
\item \emph{Refinement}: After an analysis of process performance, further
work in the analysis and design phases of the BPM cycle might be necessary. 
\end{enumerate}
In the context of these six stages, the assessment of process readiness
conducted in this paper falls into the second stage. Despite this,
we hope that the ideas and thoughts brought forward in this paper
could lead to a more formalized approach to improving macroprudential
oversight (stage one: planning). Likewise, we have not made an attempt
at improving design and modeling (stage three), as we are not presenting
a detailed documentation of the process using a standard annotation
for representing all the activities therein. With this study, our
focus is thus on establishing whether there is a common understanding
of the process, explicit or implicit, among the European System of
Financial Supervisors, and on identifying potential bottlenecks in
the underlying enablers of a mature process.

\subsection{Process Readiness}

Process analysis, as one of the six distinct phases of BPM, aims at
understanding current states of organizational processes. Given the
common complexity of today\textquoteright s processes, analysis invokes
structured approaches to gain a better understanding of the inner
workings of processes and their associated performance targets. One
widely known structured approach to understanding the state of a process
is the assessment of process readiness or maturity, as introduced
by \citet{Hammer2007,Hammer2010}.

To define a mature, high-performance process, one can assess process-specific
characteristics.\citet{Hammer2007} defines five critical enablers
for well-functioning processes: (\emph{i}) process design, (\emph{ii})
process metrics, (\emph{iii}) process performers, (\emph{iv}) process
infrastructure, and (\emph{v}) process owners. The design of the process
goes back to the specification of what tasks are to be performed,
by whom, and what the associated inputs and outputs are. This also
involves an understanding of the organization(s) involved and how
the activities interact to produce a desired outcome. The metrics
of the process detail the desired outcome in more specific terms.
Performance needs to be monitored against targets such as speed or
quality. With process design and metrics at hand, process performance
can be simulated (in relation to, for example, time, cost, and resources
needed), allowing for a better understanding of potential improvement
areas. This approach has been taken in many different contexts, such
as hospitals and banks \citep{IslamAhmed,ShimKumar2010}. The performers
of the process are those that perform the process activities. The
persons involved should not only be well versed in their particular
activity or part of the process, but also be aware of the end-to-end
design and metrics of the whole process. A suitable infrastructure
also needs to be in place for the process performers to be able to
conduct their work. Typically, this involves IT systems with the appropriate
access to information they need in order to accomplish their task.
For instance, a process improvement effort at a Swiss bank revealed
many improvement areas that were specifically related to the IT systems
that were used to support the process (\citet{KungHagen2007}). Last
but not least, a process owner is to be nominated. This is an executive
level person or organization that has the authority to oversee the
process as a whole. Without a process owner, improvements and changes
that span across different involved organizations becomes difficult
or impossible. 

The primary aim with this study is to conduct an assessment of process
readiness according to the enablers outlined above. For this purpose,
we rely on a more detailed break-down of the five process enablers,
adapted from \citet{Hammer2007}. We have designed a set of questions
to assess the maturity of each of the five enablers with their sub-categories
(all detailed in Section 4.2). Through the assessment of process readiness,
we can identify areas of improvement and set the basis for a functioning
process, also creating the basis for full-fledged process management,
if so desired. Yet, we acknowledge that the nature of processes in
macroprudential oversight, as well as some processes in the public
sector at large, is somewhat different to a standard business process.
Can these be studied with tools from BPM?

\subsection{Public collaborative processes }

Beyond profit-maximizing businesses, societies are organized around
public goals regarding the well-being of their citizens. As noted
by \citet{Olson1965}, among many others, the need for public involvement
is mostly related to markets with positive and negative externalities
and characteristics of non-excludable and non-rivalrous goods and
services, as well as fundamental human rights.%
\footnote{For a further discussion on theories of economic regulation, readers
are directed to the seminal papers by \citet{Stigler1971} and \citet{Posner1974},
as well as a later review by \citet{Peltzman1989}.%
} In addition to involvement only in minor market segments, we define
a public goal as a broader ultimate objective provided by the government
to the population within its jurisdiction, which might require involvement
of a set of public, private and third sector actors.

Comprising of multiple providers acting together for a common purpose,
the actors involved in public goals are oftentimes dispersed. Examples
of these societal goals include public safety, education and health,
as well as a well-functioning financial system. Oftentimes, we also
see actors taking on roles that are typically not associated with
their primary goal. One such example could be schools educating children
in traffic rules and fire hazards, and thus promoting public safety.
The police might be offering lectures on drug abuse, and thus promoting
public health. Thus, public institutions work together to build a
well-functioning society, oftentimes taking on different roles depending
on an individual\textquoteright s particular needs. Sickness, criminal
behavior, age, disabilities, and personal interests are all relevant
examples of conditions that can trigger a particular response from
society. In some instances, the response is highly regulated and clearly
defined, such as in the case of criminal misconduct, but oftentimes
there is a high degree of informal co-operation and implicit understanding
of how to promote a particular goal. The case of public health is
such an example. If we disregard the most obvious example of an actor
in this field, hospitals, which are usually involved when a de facto
health problem already exists, we see a host of actors involved in
preventive measures to avoid health issues. This is highly dependent
on specific structures in various countries, but it is not uncommon
to see a diverse set of actors in fields like education, food safety,
and sports promoting health, in addition to the explicit healthcare
system. As such, many different actors are then involved in the process
of ensuring public health.

We denote the activities performed by various societal actors working
together for a common purpose as a public collaborative process (PCP).
This is defined as a set of public, private, and third sector actors,
working together in a collaborative manner for a societal goal, which
is defined and possibly regulated at a governmental level. The complexity
of these types of processes becomes apparent when one moves from a
high-level description to a lower level, where responsibilities and
activities performed by individual organizations are disentangled.
The high-level goals for PCPs are typically explicitly documented,
and there is often a primary organizational entity that assumes the
overall responsibility for the process. However, at a lower level,
the activities and cooperation performed by a dispersed set of actors
are oftentimes more unstructured in nature, such as following less
formal documentation and being more dependent on bilateral agreements.
Mostly, these activities are also coordinated at a high level, such
as by a responsible ministry, potentially creating a gap between the
operative and governing actors.

\begin{figure}[h]
\begin{centering}
\includegraphics[bb=0bp 0bp 934bp 541bp,width=0.9\columnwidth]{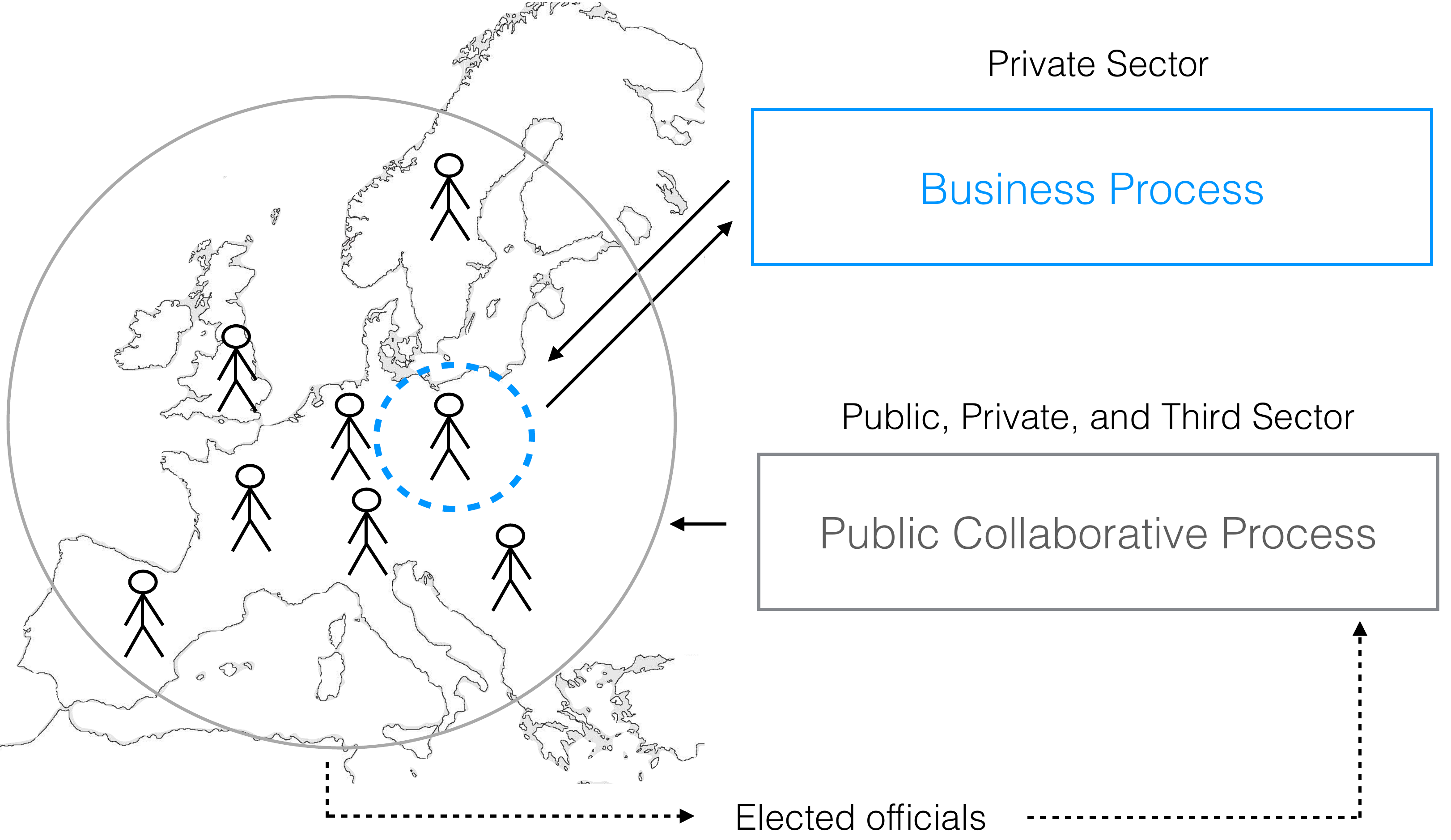}
\par\end{centering}

\textbf{\scriptsize{}Notes}{\scriptsize{}: The figure represents a
comparison of PCPs and standard business processes. While business
processes are shown to target chosen, individual customer segments,
PCPs serve entire populations. Likewise, the figure shows that the
feedback loop of PCPs through elected officials is complex in nature,
whereas in standard business processes customers choose whom to buy
from.}{\scriptsize \par}

\centering{}\protect\caption{\label{fig:Public-collaborative-processes}Public collaborative processes
and business processes}
\end{figure}
Macroprudential oversight is clearly a PCP. To begin with, its goal
of financial stability can be seen as a public good in that the producer
cannot exclude anybody from consuming it (i.e., non-excludable) and
consumption by one does not impact that of others (i.e., non-rivalness).
Further, it involves collaboration of a large number of dispersed
actors to reach the common goal, and their cooperation is fairly unstructured.
A crucial question to ask is whether PCPs, including macroprudential
oversight, can and should be managed with the same means as business
processes. We believe they should, since they contain the same elements.
The basic function of any process is to transform inputs to outputs
with the ultimate goal of serving customers. This could take on the
form of delivering a product or a service, ensuring that a curriculum
is taught and understood, ensuring that a particular illness is prevented
or treated, or ensuring that financial stability is safeguarded to
enable a well-functioning financial system. This basic and fundamental
element is common for both business processes and PCPs. The justification
of any process, public or private, is that it serves its ultimate
customers.

There are, however, differences between business processes and PCPs,
specifically in how processes are adapted to serve customers. Figure
\ref{fig:Public-collaborative-processes} illustrates two challenges:
indirect and complex feedback loops. A crucial element in firm competitiveness
is adaptability to customer needs (e.g., \citet{Porter1985}). With
the same token, any PCP should ultimately be adapted to serve its
customers, the citizens of any given country. However, the feedback
loop between the provider and the customer differs substantially when
comparing PCPs to business processes. While the firm typically has
a rather direct link to their customers, the same cannot be said of
PCPs. Instead, any fundamental change to the structure, performers,
and function of PCPs is regulated through an oftentimes democratic
system, with politicians serving as the representative of the customer,
the people. Accordingly, another dimension of complexity derives from
the fact that PCPs need not only serve a target group of customers,
but rather the population in its entirety. 

Despite these challenges may explain the characteristics of PCPs,
such as a dispersed set of actors cooperating under unstructured forms,
we see no reason why PCPs would not be examined through the means
of process management. Accordingly, the lengthy and complex feedback
loop in PCPs does not reduce the need to formalize processes. Beyond
a more efficient and effective PCP, low process readiness only serves
to create confusion, whereas well-documented and understood PCPs serve
to direct the attention of both voters and elected officials to matters
of importance.

\section{Assessing the maturity of macroprudential oversight in Europe}

This section presents the overall research design, including approaches
for data collection and methodological perspectives, the analysis
model used as a basis for drawing conclusions of the data, and discussion
and analysis of the collected data with the help of the model.

\subsection{Research design }

For the purposes of this study, we have conducted a series of interviews
involving actors that operate within the European System of Financial
Supervisors (ESFS). Initially, we started out by conducting four semi-structured
interviews. The aim with these interviews was to assess a suitable
level of analysis, that is, how should we go about looking at macroprudential
oversight with the lens of process management. Our interpretation
of the semi-structured interviews was that macroprudential oversight
is not mature enough for full process management. Although there was
an agreement of the aim to effectively and efficiently safeguard financial
stability, process design and modeling (with subsequent steps in the
six BPM phases) seemed elusive at this stage. Hence, we decided to
approach process management from a more elementary perspective by
assessing process readiness.

We continued data gathering through interviews. At this stage, we
opted for a more structured approach with pre-defined questions that
relate directly to the five process enablers. The questions are detailed
in Table A.1 in the Appendix. In total, we conducted 23 additional
interviews, of which four are written answers from the four interviewees
of the semi-structured interviews. This involved a total of 12 organizations
within the ESFS. The interviews lasted around 75 minutes each, except
for the four written accounts with answers to the same questions.
All of the interviews were conducted within May and June, 2014.

Our aim was to have a broad coverage of the organizations depicted
in Figure \ref{fig:ESFS}. This involves covering both national and
European organizations, and policymakers and supervisors, as well
as actors within and outside central banks, and with and without a
direct macroprudential mandate. In an effort to avoid elite bias (involving
only managerial viewpoints), we wanted to ensure not only a broad
organizational coverage, but also insights to the opinions of people
on all levels within the involved organizations. Generally, we aimed
at approaching two experts within each involved organization, out
of which one was an operative expert and one a governing expert, both
from financial stability functions. Governing experts refer to policymakers
and supervisors involved in heading or managing a division or department
with financial stability responsibility, whereas operative experts
refer to analysts and economists involved in data management and analysis
in a similar division.

\begin{figure}[H]
\begin{centering}
\includegraphics[width=0.6\columnwidth]{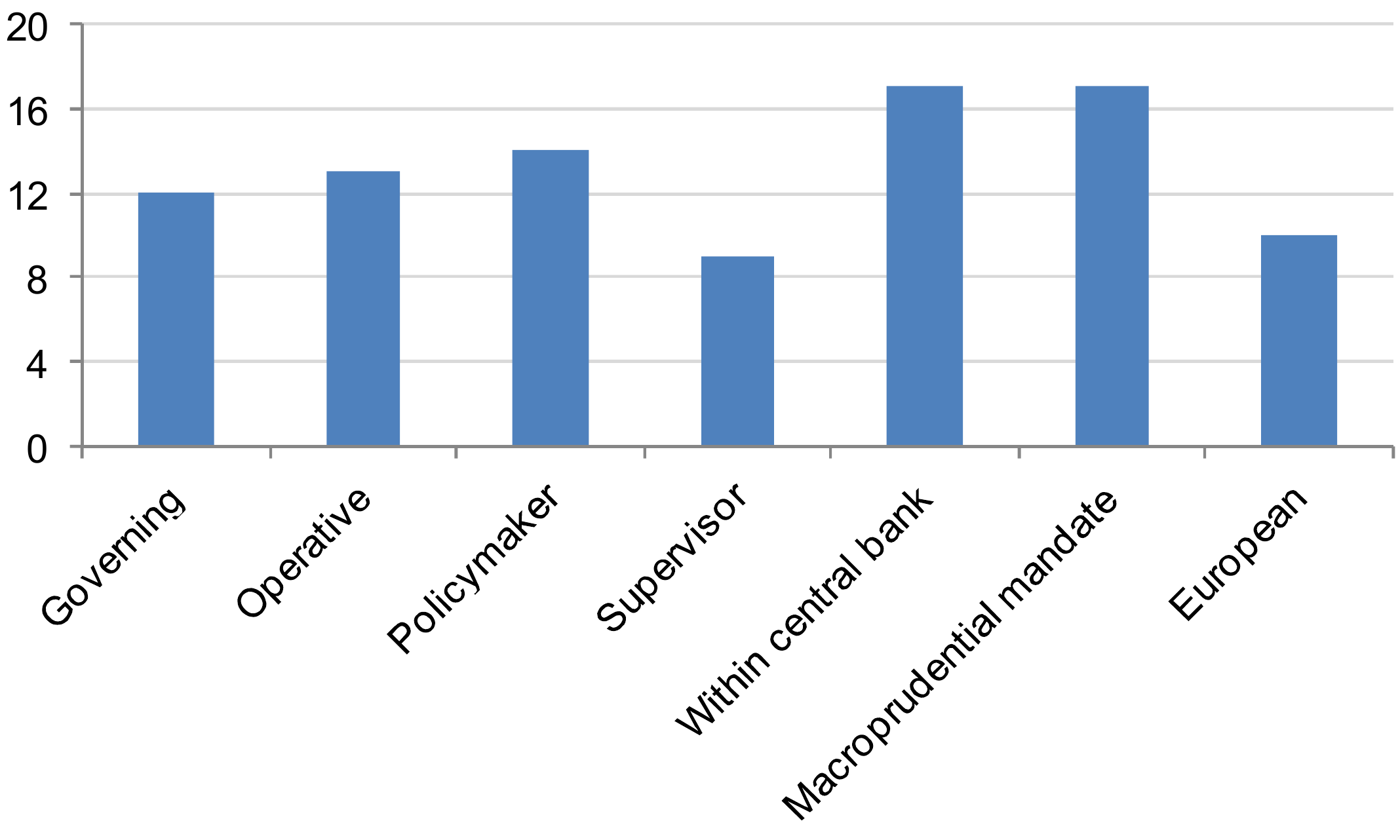}
\par\end{centering}

\textbf{\scriptsize{}Notes}{\scriptsize{}: This bar chart shows statistics
of the interviewee sample's characteristics. For each property, a
bar shows the number of individual respondents that fulfill that specific
characteristic.}{\scriptsize \par}

\centering{}\protect\caption{\label{fig:Demographics-of-interviewees}Characteristics of the interviewees}
\end{figure}

In Figure \ref{fig:Demographics-of-interviewees}, we present statistics
of the interviewee sample\textquoteright s characteristics. Out of
23 interviewees, 12 were classified as governing and 13 as operative
experts. The sum does not add up as expertise is not mutually exclusive.
Due to the small size of their organization, particularly departments
focusing on macroprudential tasks, two interviewees were assessed
to be in charge of both governing and operative tasks. In a mutually
exclusive classification of interviewee\textquoteright s roles, our
sample consists of 14 policymakers and 9 supervisors. The division
into policymakers and supervisors follows to a large extent the division
between microprudential and macroprudential organizations, with exceptions
due to variation in institutional models. For instance, supervisory
authorities may be mandated with macroprudential tasks or may be located
within the central bank. Hence, we also report that 17 of our interviewees
are in an organization with a macroprudential mandate and the same
number are located within a central bank. Distinguishing between national
and supranational organizations, our sample consists of 10 interviewees
at a European level and 13 at a national level.

To promote an open discussion, we give interviewees full anonymity.
Furthermore, due to the sensitive nature of the topic, we also made
a decision not to record the interviews. With these measures, we feel
that there were few inhibitions with regard to our interviewees expressing
their true opinion. During the interviews, a brief summary of each
answer was written down as the interview progressed. For our data
analysis, we relied on this written documentation to obtain an overview
of the opinions of our interviewees. Each answer was given a score
between 1 and 5, one representing a low readiness with regard to the
process enabler being measured, and five representing a high readiness.
The scores were given based on a joint discussion between the two
authors of this paper. In Table A.2 in the Appendix, we provide guidance
on our scaling through examples of given scores. Furthermore, for
each measured process enabler and its subdimensions, we also aim at
obtaining a measure of consensus between our interviewees through
the variation in scores given by us.

\subsection{The analysis model}

\begin{figure}[H]
\begin{centering}
\includegraphics[width=0.9\columnwidth]{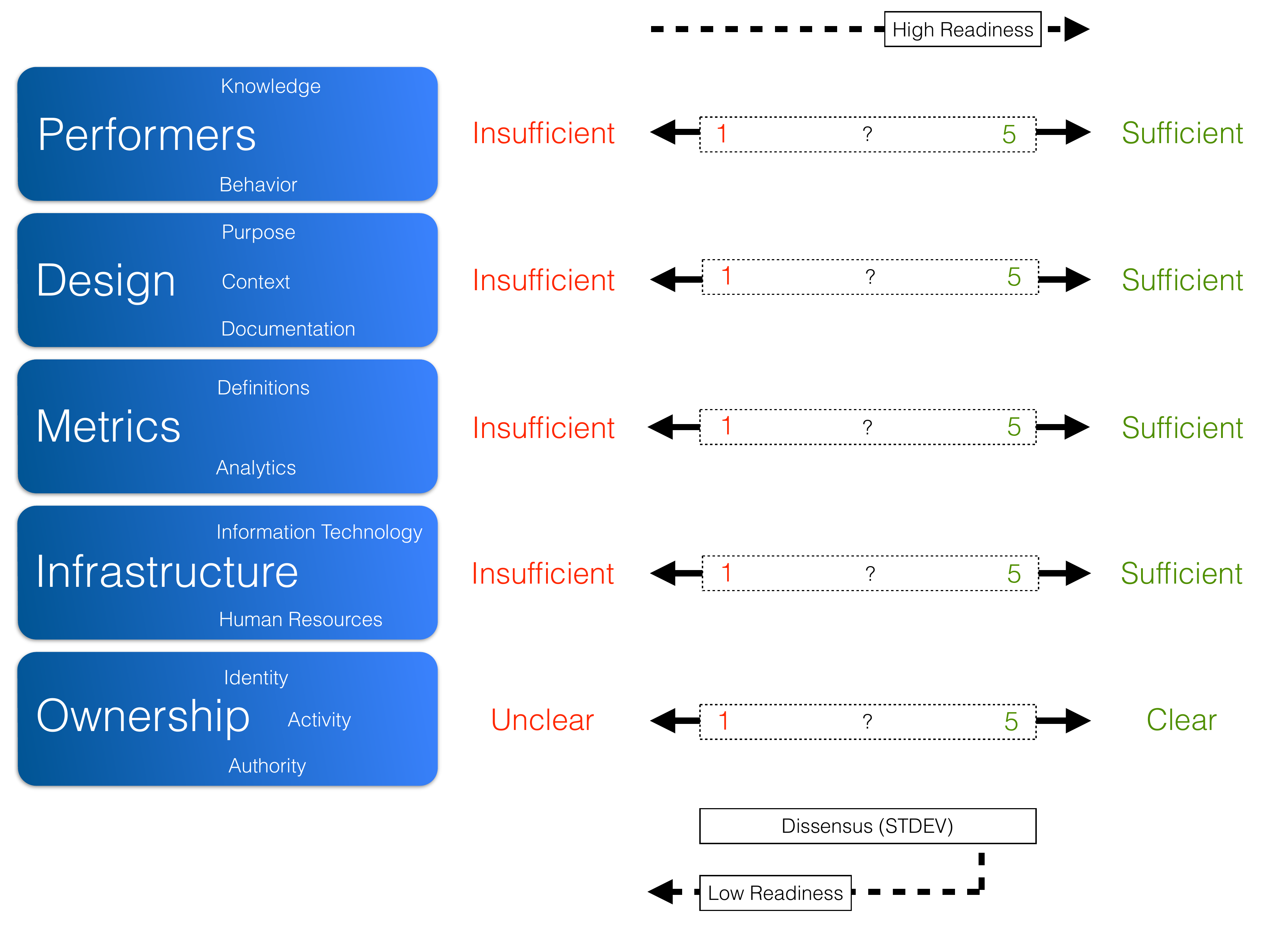}
\par\end{centering}

\textbf{\scriptsize{}Notes}{\scriptsize{}: The figure presents the
analysis model that is put forward in this paper. For each measured
process enabler (and its subdimensions), we represent two measures:
level and dissensus of readiness. Level of readiness is represented
through positions of a marker along the horizontal dimension, whereas
dissensus is shown through the size of the rectangular marker.}{\scriptsize \par}

\centering{}\protect\caption{\label{fig:The-analysis-model}The analysis model}
\end{figure}

This section presents the analysis model used to assess the macroprudential
oversight process. First, we describe how we mapped the interview
questions to the process enablers discussed in Section 3.2. Then,
we provide a framework for assessing process maturity of macroprudential
oversight, in which we measure both the level and dissensus in process
readiness. Figure \ref{fig:The-analysis-model} presents the analysis
model that we put forward herein.

Following the set-up in Section 3.2, we describe process readiness
through the five process enablers: process performers, design, metrics,
infrastructure and owner. In order to fit this framework to the process
of macroprudential oversight in Europe, we customized our interview
questions to the process enablers. The questions aim at capturing
the level of readiness for each enabler; compared to Hammer\textquoteright s
\citep{Hammer2007} original framework we have reformulated the questions
to fit the domain in question. At the beginning of each interview,
we stressed that we wish to have a European focus throughout the conversation.
While the precise formulation of the questions can be found in Table
A.1 and the scaling of our scores in A.2, both in the Appendix, the
below discussion focuses on the enablers and their subdimensions that
we aim at measuring.
\begin{itemize}
\item First, we assess process performers through the skillset and mindset
of involved actors. Knowledge is assessed through familiarity with
macroprudential oversight, including both issues related to policy
and decisionmaking, and analysis and assessment of financial stability.
Behavior focuses on intrinsic motivation and true interest in improving
and developing macroprudential oversight. 
\item Second, the assessment of process design focuses mainly on a step-by-step
specification, or end-to-end design, of macroprudential oversight.
This is disentangled into three subdimensions: (\emph{i}) purpose,
a clear end-to-end understanding of macroprudential oversight; (\emph{ii})
context, an awareness of inputs and outputs among institutions; and
(\emph{iii}) documentation, a clear and accessible specification of
tasks to be performed. 
\item Third, we assess process metrics through the overall use of measurement
in steering macroprudential oversight as a process. This mainly concerns
the definitions and uses of metrics to ensure efficiency and effectiveness
and a balance between costs and benefits in macroprudential oversight,
but also involves the use of analytics in measuring systemic risk. 
\item Fourth, we assess the extent to which tools and human resources provide
a sufficient infrastructure for supporting tasks in macroprudential
oversight. Tools refers broadly to infrastructure provided by available
policy interventions and IT and data-related support functions. Human
resources as an infrastructure focuses on the sufficiency of hiring,
training and development as a support for macroprudential functions. 
\item Fifth, the assessment of process owners focuses on responsibilities
and mandates to oversee macroprudential oversight. More specifically,
we are concerned with the responsible entity for well-functioning
macroprudential oversight, for guiding development activities and
with ultimate mandate to implement changes.
\end{itemize}

We view these five process enablers through an analysis model that
measures the level of readiness and lack of consensus (or dissensus)
for each enabler. Beyond quantifying individual interviewees\textquoteright{}
responses into a scale between 1 and 5 for each process enabler, as
well as their subdimensions, we also measure the level of readiness
and dissensus through quantitative means. The level of readiness $L$
is computed through a simple arithmetic mean of the observed values
$a_{1},a_{2},...,a_{n}$:

\[
L=\frac{1}{N}\sum_{i=1}^{N}a_{i}
\]

where $N$ stands for the size of the sample. The dissensus in readiness
$D$ of an enabler is measured through the standard uncorrected sample
standard deviation:

\[
D=\sqrt{\frac{1}{N}\sum_{i=1}^{N}(a_{i}-L)^{2}}
\]

where $a_{1},a_{2},...,a_{n}$ are again the observed values and $L$
is the mean value of the observations. We maintain that a low consensus,
representing inconsistency in the scoring of a given process enabler,
is indicative of an unclear overall status, thus lowering the readiness
for that enable. In Figure \ref{fig:The-analysis-model}, each process
enabler is broken down into a number of subdimensions, for which the
level of readiness is represented with the position of a rectangular
marker on the horizontal axis (higher readiness to the right) and
dissensus as the size of the rectangle.

When mapping the questions to the level of readiness, it should be
noted that for some questions the explicit answer is less interesting
than the ability to provide an answer in the first place. Such is
the case, for example, when we inquire about improvement possibilities
to macroprudential oversight. We view clear improvement suggestions
as a strong willingness to improve the process, and are less interested
in the actual improvement proposal, albeit also important in itself.
Moreover, when capturing lack of consensus (or dissensus) in terms
of variation in the assessed readiness levels, it is also worth noting
that the measure as such is not always reflecting actual dissensus.
For instance, despite the interviewees agree upon the existence of
process owners, and can name them, it does not reflect the fact that
they disagree on the identity of the actual owner.

\subsection{Discussion }

\begin{figure}[H]
\begin{centering}
\includegraphics[width=0.8\columnwidth]{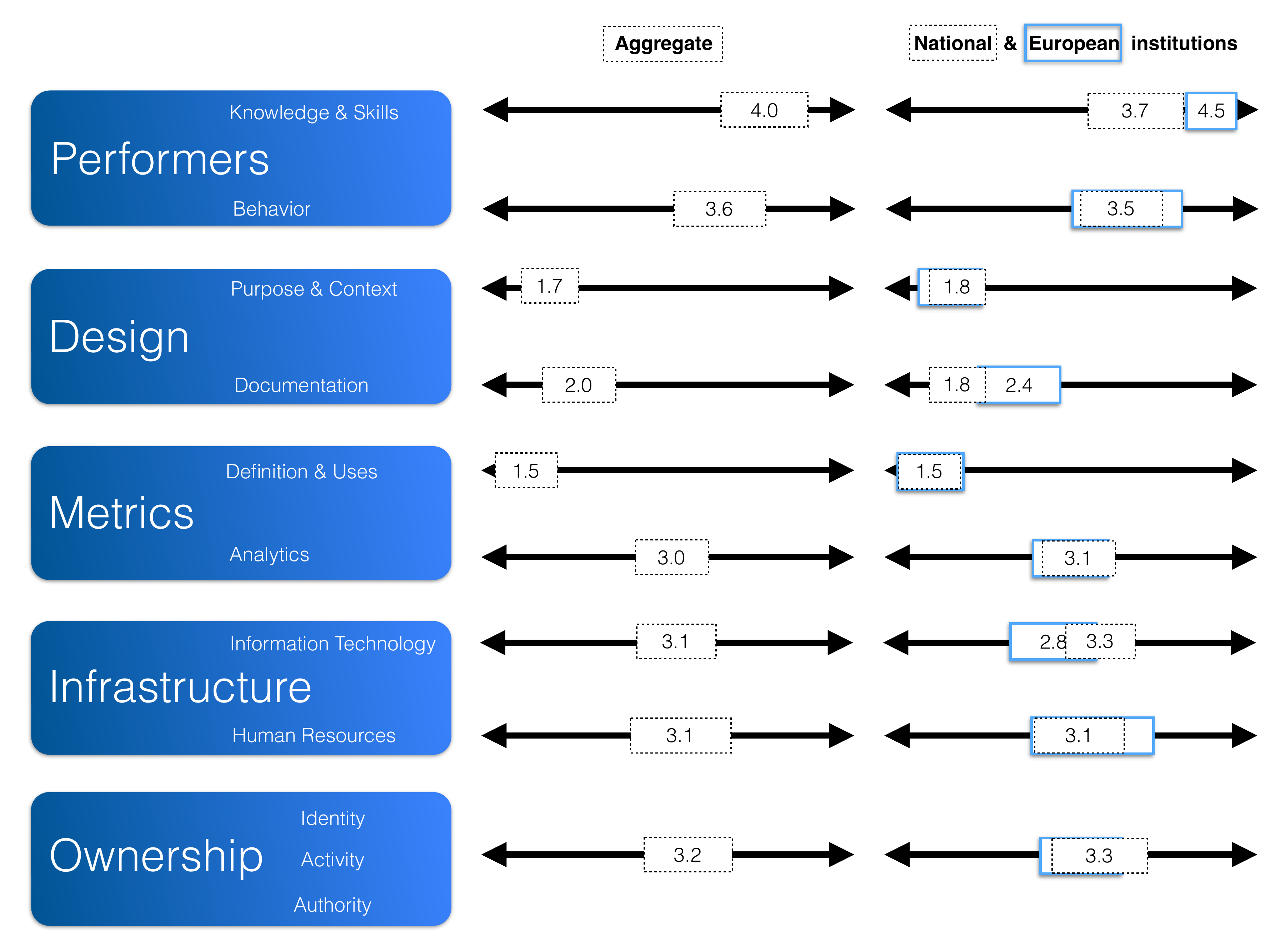}
\par\end{centering}

\textbf{\scriptsize{}Notes}{\scriptsize{}: The figure presents an
assessment of the process of macroprudential oversight in Europe.
Level of readiness is represented through positions of a marker along
the horizontal dimension, whereas dissensus is shown through the size
of the rectangular marker. The figure represents on the left an aggregate
result and on the right a result disaggregated for respondents in
national and European institutions, respectively.}{\scriptsize \par}

\centering{}\protect\caption{\label{fig:Assessing-the-process}Assessing the process of macroprudential
oversight in Europe}
\end{figure}

With the above presented research and analysis model as a basis, this
section discusses the results of our empirical analysis. Measuring
the level of and dissensus in process readiness in our sample allows
us to assess the maturity of macroprudential oversight in Europe.
We present the results in Figure \ref{fig:Assessing-the-process}.
Broadly, our analysis shows the following observations. To start with,
we observe that \textquoteleft process design\textquoteright{} (including
purpose, context, and documentation) and \textquoteleft process metrics\textquoteright{}
(particularly definitions and use of metrics) exhibit the lowest levels
of readiness. The level of consensus is high, or in other words, there
seems to be an agreement with regard to the challenges in this area.
The \textquoteleft process performers\textquoteright{} enabler exhibits
a relatively high level of readiness but with some disagreement in
comparison to design. Analytics to steer process metrics, as well
as the \textquoteleft process infrastructure\textquoteright{} and
\textquoteleft process ownership\textquoteright{} enablers, exhibit
an average level of readiness, but with larger dissensus. In particular,
despite agreement on the existence of a process owner, the lack of
consensus on the ownership is especially alarming due to the range
of suggestions. Beyond these numerical aggregates, but with them as
a starting point, we also tap into the richness of the underlying
interview data in the following.

The \emph{performers} as a process enabler is generally seen as sufficient.
Overall, our respondents are familiar with the current setting in
macroprudential oversight, in particular from the perspective of their
own task, but also from an end-to-end perspective. Most interviewees
can name the involved actors. Yet, it is clear, and also obvious,
that the ongoing development of macroprudential oversight causes uncertainty
due to the lack of visibility to the future. It is also worth noting
that our interpretation of questions related to this enabler is indirect
in nature, as to minimize a potential sampling bias due to all interviewees'
having a financial stability responsibility. There are certain discrepancies
in the answers to the questions related to the behavior of performers.
While this can partly be explained by different roles and responsibilities
(national vs. European level and hierarchical position), there are
some who show a degree of cynicism as to their opportunities to influence
and the overall chance of reaching well-functioning macroprudential
oversight. As above, the lack of clear insight to the future might
have introduced a bias, lowering the perception of the overall willingness
to improve the process at a European level.

At a general level, we can clearly observe a lack of readiness in
\emph{process design}. Given the risk of over-emphasizing the political
and legislative context in which macroprudential oversight operates,
we designed the questions to have a focus on the internal process
(i.e., the tasks as described in Figure \ref{fig:The-macroprudential-oversight}).
Despite this, the discussion often revolved around legal and political
challenges. These aspects include challenges when moving between the
national and European level, challenges that are further accentuated
when moving outside states in the euro area. Overall, a great deal
of confusion exists regarding who is responsible for different parts
of macroprudential oversight. From a context perspective, discussions
have highlighted more conflicts between various processes (i.e., policy
objectives) than synergies. Further, no documented end-to-end process
exists. Many interviewees referred to legislation, although these
are not directly comparable to work descriptions. National level seems
to be better documented and more easily accessible than European level.

In terms of \emph{process metrics}, despite wide agreement on the
need for them, there are currently no measures of well-functioning
macroprudential oversight. In certain cases, references were made
to measures of systemic risk, even though there is agreement that
they are not the same as well-functioning macroprudential oversight.
As a consequence, it is impossible to say whether costs and benefits
of macroprudential oversight is aligned, something that some interviewees
pointed out would be crucial in gaining customer justification (i.e.,
acceptance from the public). Analytical models do play a key role
in macroprudential oversight, which is agreed upon by almost all interviewees.
Despite wide agreement on the use of analytics, the interviewees agree
equally much upon the fact that systemic risk is challenging to quantitatively
measure. Even with the right metrics in place, measurability and the
changing nature of risk have been highlighted as issues, such as concerns
with shadow banking.

\emph{Infrastructure} for macroprudential oversight is generally seen
as work in progress; there seems to be agreement on differences in
readiness. Policy instruments and tools are available, although possibly
still requiring development and amendments in legislation to allow
for implementation of suitable measures. IT infrastructure seems to
be sufficient but to the contrary, there seems to be wide agreement
on needed improvements to the overall data infrastructure, particularly
dealing with big data, European-level databases, data quality checks,
cross-sectional comparability and linking various sources. One general
challenge seems to be related to data sharing, ranging from turf issues
to legal issues and lags in making it available. There is consensus
on a sufficient overall human resources infrastructure, which supports
in hiring of employees. However, interviewees voiced concerns related
to the availability of people who have knowledge and know-how of macroprudential
oversight. As the field is new, there is a lack of people with the
necessary skills that would directly support the tasks at hand. This
also created a large discrepancy in the answers, and an increase in
dissensus, depending on the focus of the respondent. According to
some, but not all, training and development of existing resources
is available, to the extent that time permits. In our view, the level
of existing training depends also on the maturity of other process
enablers, such as end-to-end design, measures of financial stability
and analytics.

Regarding \emph{ownership}, many respondents mentioned one or more
responsible entities as being in charge of well-functioning macroprudential
oversight (including the development of the process, and overseeing
tasks and performers). At the same time, many interviewees acknowledged
challenges in this area, clearly indicating that they do not know
who is in charge. Furthermore, the range of actors mentioned by those
that were willing to point at the entity or entities in charge was
very broad.%
\footnote{To exemplify the wide range of the mentioned involved actors, the
interviewees named the following bodies or persons: ESRB, FSC, FSB,
European Council, Heads of State, ECB, ECB's Governing Council, Mario
Draghi, ESFS, European Commission, Voting citizens, European Parliament,
SSM (as of Nov 2014), ESRB\textquoteright s General Board, national
authorities including central banks (jointly and individually mentioned
by respondents), Ministries of Finance, Eurogroup Working Group, ECOFIN,
\emph{\textquotedblleft those responsible for policy implementation\textquotedblright{}}
and \emph{\textquotedblleft ECB for euro area and ESRB for Europe\textquotedblright },
as well as \emph{\textquotedblleft no one and all\textquotedblright }. %
} It would seem many organizations are involved but no one is having
the overall responsibility for the process. Our approach to scoring
indicated fairly low dissensus, but this does not capture the differences
in the named responsible actors. As such, there is little consensus
on individual process owners.

Beyond the level of and dissensus in readiness for individual process
enablers, one should not forget that the enablers are mutually interdependent.
In case any one enabler is missing or insufficient, the rest of the
enablers are prone to be ineffective. \citet{Hammer2007} exemplifies
interdependence as follows: \emph{``A weak owner can't implement
a strong process design, poorly trained performers can't carry out
the design, a bad design can't optimize the process metrics {[}...{]}.''
}In the context under analysis, we can observe a number of interdependency-linked
challenges related to the enablers with lowest readiness: process
design, metrics and ownership. Starting out from improvements to the
process design, one would need other process enablers to guide and
implement changes. While process metrics ought to provide guidance
on bottlenecks in the process design, the ultimate implementation
of changes is to be carried through by the process owner. These can
be seen as clear hinders to process design improvements. Likewise,
as the definitions and uses of metrics show the lowest readiness,
and respondents struggle in defining systemic risk, it is neither
clear nor convincing that systemic risk analytics can be carried out
effectively, despite its average readiness. In addition, we have observed
that weaknesses in process enablers may also derive from others enablers,
such as challenges in data infrastructure (process metrics) being
a process design problem related to data sharing. 

In addition to the five process enablers, the interviewees were asked
a final question on the key challenge and how they would improve macroprudential
oversight in Europe.%
\footnote{The wording of the question was as follows: \emph{``Forget all restrictions:
How should macroprudential oversight in Europe be improved? What would
it take for this to happen?''}%
} The aim of the question was to highlight the most significant challenges
and to enable a discussion of issues outside the scope of previous
questions. In the following, we will describe a summary of the discussed
issues. The most prominent challenge related to the lack of clear
mandates and responsibilities, and the complexity of the system of
involved actors. A number of respondents highlighted many overlapping
layers in actors' tasks and the need to streamline procedures, structures
and the overall design of the process. In this context, many also
highlighted problems with data availability and sharing, again pointing
to these issues being a design problem. Relating to metrics, respondents
also requested further research and guidance on the following issues:
definitions of systemic risk, how financial instability happens and
the link between policy objectives and tools. Moreover, interviewees
also pointed out that we are in the beginning of macroprudential oversight
in Europe, and that time will show and guide us in how things ought
to develop. In fact, interviewees even hinted value in the type of
assessment proposed and conducted in this paper.

A final view to the analysis model descends from various aggregations.
Instead of considering all respondents, we can choose to focus on
subsets based upon their roles (see Figure \ref{fig:Demographics-of-interviewees}).
The respondents may be categorized based upon institutional models
(see Figure 2) and policy mandates (see Figure 3), as well as differences
in the profiles of actors (see \citet{Goodhartetal2002} for a comparison
of policymakers and supervisors). This not only provides insights
into process readiness, but can also be seen as a type of robustness
check with respect to the sampling of respondents. The left column
in Figure \ref{fig:Assessing-the-process} provides results disaggregated
with respect to respondents from national and European institutions.
Generally, we can observe that the results are similar in nature.
When assessing the largest differences, we can observe that European
actors perceive themselves to be more knowledgeable about macroprudential
oversight and assess the documentation to be of better quality. This
is most likely in line with the fact that actors at the European level
are in a position to have better (and potentially earlier) insights
into European macroprudential oversight, and might thus also be better
aware of the prevailing documentation. The lower level of readiness
in IT infrastructure might not only point to challenges in these issues
in European organizations, but might potentially also highlight that
information systems and data warehouses are more complex when dealing
with pan-European infrastructure. The dissensus is higher for all
measures, except process performers, which highlights the heterogeneity
among European actors. Following all the rest of the characteristics
of the respondents presented in Figure \ref{fig:Demographics-of-interviewees},
we report disaggregated measures in Table A.3 in the Appendix. Despite
slight variation depending on the assessed aggregation, the table
confirms robustness of the above discussed results in that the same
conclusions hold.

\section{Conclusion}

This paper has illustrated a process perspective to assessing macroprudential
oversight in Europe. As a sequence of activities with the ultimate
aim of preventing and mitigating systemic risk, macroprudential oversight
can be viewed as an inherently complex process, not the least the
European System of Financial Supervisors with its large number of
actors at national and supranational level. To conceptualize a process
in this context, we introduced the notion of a public collaborative
process (PCP). PCPs involve multiple organizations with a common objective,
where the dispersed organizations cooperate under various unstructured
forms and take a collaborative approach to reaching the final goal.
In this paper, we have argued that PCPs can and should be managed
using the tools and practices common for business processes.

At a more general level, the absence of well-functioning, transparent,
and documented PCPs only act to support populism and simplified solutions.
If the problem is not known, solutions are hard to find. As such,
process readiness in the context of PCPs serves to enhance the political
system, a well-functioning democracy. The globally upcoming objective(s)
of macroprudential oversight is neither a simple task to tackle nor
a dimension free of politics. As the European set-up is in the making,
now is the time to understand activities, define responsible entities
and discern how the actors interact, which we propose to be done through
the lens of process management.

To analyze the macroprudential oversight process, we have conducted
an assessment of process readiness through interviews with actors
in European macroprudential oversight. Based upon the interviews,
we provided an analysis model to assess the maturity of five process
enablers for macroprudential oversight. Broadly, when measuring the
level of and dissensus in process readiness of macroprudential oversight
in Europe, our analysis shows the following observations. To start
with, we observe that process design and metrics exhibit the lowest
levels of readiness. For both, the level of consensus is high. The
enabler of process performers exhibits a relatively high level of
readiness but with some disagreement in comparison to design. Analytics
to steer process metrics, as well as the process infrastructure and
the enabler of process ownership exhibits an average level of readiness,
but with larger dissensus. Beyond these numerical aggregates, we have
also tapped into the richness of the underlying interview data.

Whereas the results of our analysis point to clear recommendations
on the areas that need further attention when macroprudential oversight
is being developed, the above concluding summary provides only a snapshot
of the maturity of the process, which is likely to be somewhat outdated
when this paper goes to press. Hence, we would like to see that we
have provided a general purpose framework for assessing process readiness,
rather than an ending point. The framework lends itself to regular
updates of the assessment of process readiness, enabling monitoring
the impact of improvement efforts over time. Likewise, this framework
is far from bound to the region under analysis in this paper, not
the least to assessments of the state of macroprudential oversight
in the US and UK.

\newpage{}

\section*{References}

\section*{\textmd{\small{}\renewcommand\refname{References}\bibliographystyle{plainnat}
\bibliography{References/references}
}}

\section*{\textmd{\small{}\newpage{}}}

\setcounter{table}{0}

\renewcommand{\thetable}{A.\arabic{table}}

\setcounter{section}{0}

\renewcommand{\thesection}{Appendix A}

\section{Questionnaire, scale and aggregations}

\begin{table}[H]
\protect\caption{The questionnaire and a mapping to process enablers.}

\noindent \centering{}\includegraphics[width=1\columnwidth]{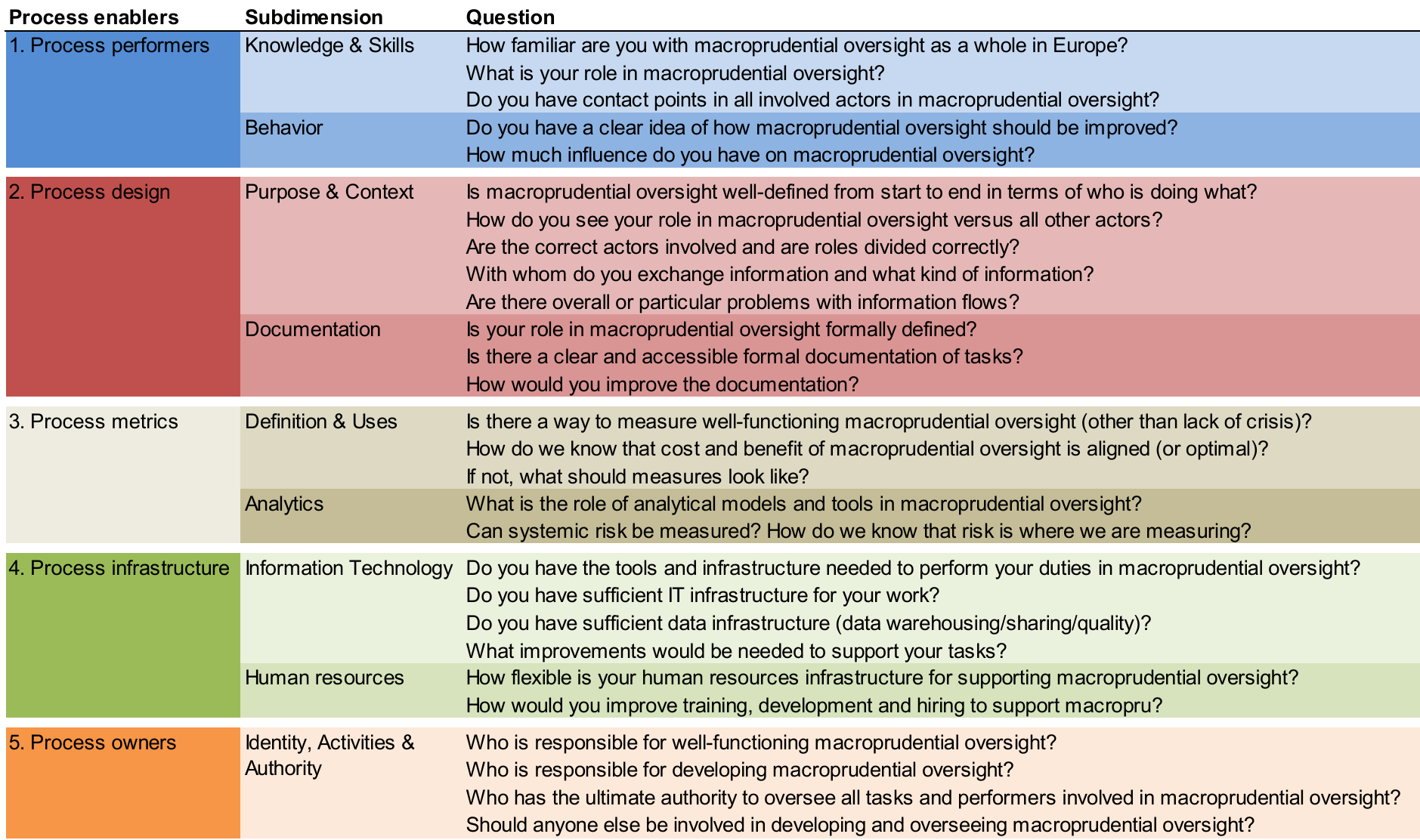}
\end{table}

\section*{\textmd{\small{}\newpage{}}}

\begin{table}[H]
\protect\caption{Examples of our scoring and scaling.}

\noindent \begin{centering}
\includegraphics[width=1\columnwidth]{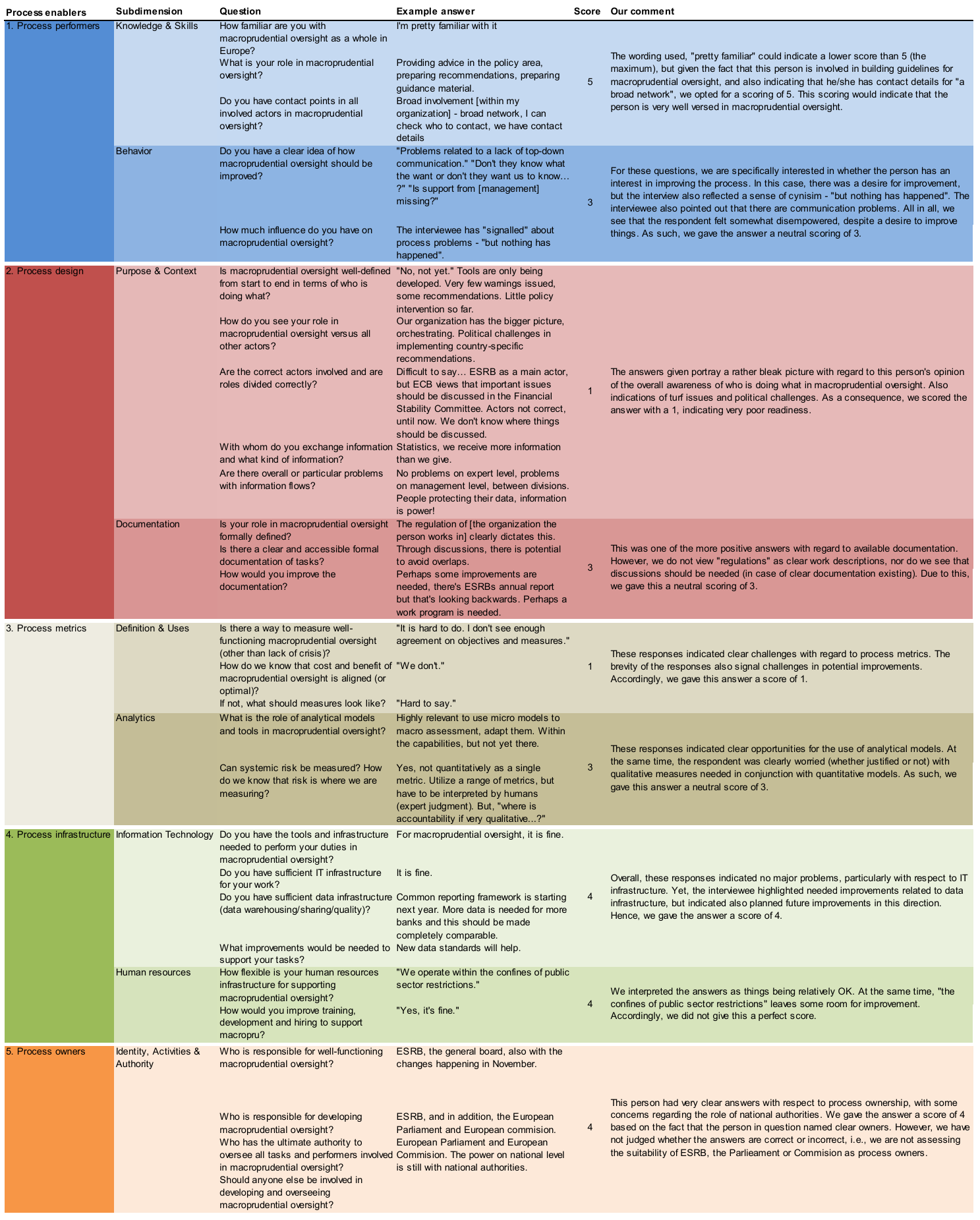}
\par\end{centering}

\textbf{\scriptsize{}Notes}{\scriptsize{}: The table shows for each
process enabler and its subdimensions an example answer, scoring and
a comment describing the reasoning behind the score.}
\end{table}
\begin{table}[H]
\protect\caption{Aggregated and disaggregated results with the analysis model.}

\noindent \begin{centering}
\includegraphics[width=1\columnwidth]{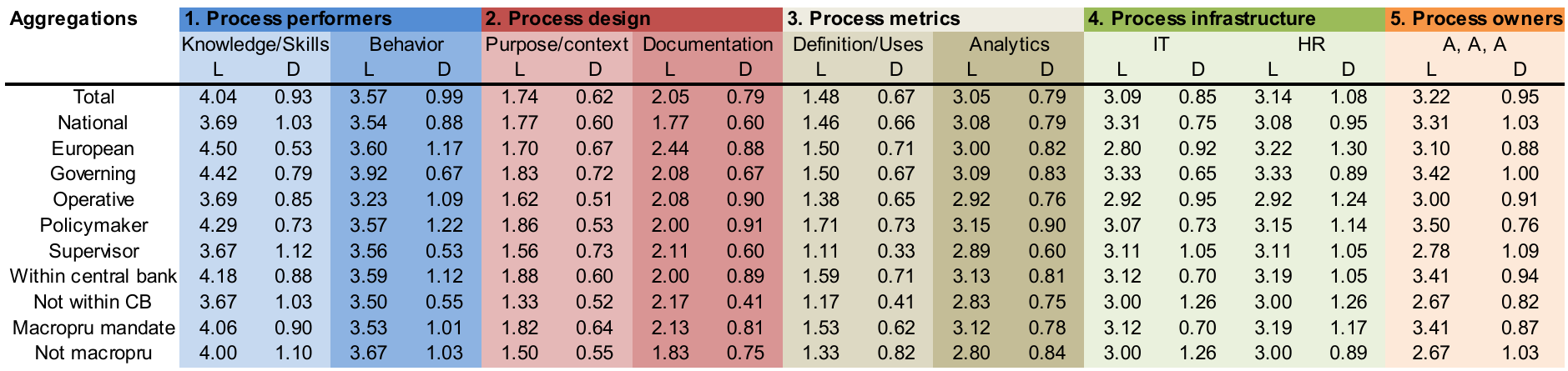}
\par\end{centering}

\textbf{\scriptsize{}Notes}{\scriptsize{}: The table reports the level
of readiness (L) and dissensus (D) for each process enabler and its
subdimensions for a number of different aggregations. The aggregations
follow the characteristics presented in Figure \ref{fig:Demographics-of-interviewees}.}
\end{table}

\end{document}